\documentclass[final,english]{bullsrsl}[2022/06/15]

\usepackage[latin1]{inputenc}
\usepackage[T1]{fontenc}

\usepackage{natbib} 
\usepackage{graphicx}
\usepackage{subfigure}

\begin{document}
\title{Formation scenarios of Ba stars: new evidence from the masses of the companion AGBs}

\author[affil={1,2}, corresponding]{Partha Pratim}{Goswami}
\author[affil={1}]{Aruna}{Goswami}
\affiliation[1]{Indian Institute of Astrophysics, Koramangala, Bangalore
    560034, India}
\affiliation[2]{Pondicherry University, R.V. Nagar, Kalapet, 
605014 Puducherry, India}
\correspondance{partha.goswami.iiap@gmail.com}
\date{19th April 2023}
\maketitle

\begin{abstract}
The abundances of the slow neutron-capture (s-) process elements observed in barium stars can be explained by considering  mass transfer in a binary system from an asymptotic giant branch (AGB) star onto a smaller mass and less evolved binary  companion star.  Based on the abundances of neutron-capture elements the barium stars are broadly divided into two categories,  ``strong''  and ``mild'' barium stars. However, an understanding of the role of the characteristic properties of the companion AGBs on the formation of the different types of barium stars is still lacking. This work focuses on investigating the role of the mass of the companion AGB stars in this context. In a recent high-resolution spectroscopic study, we studied in detail the chemical composition of a few barium stars that we have identified. In order to understand the nature and the mass of the companion stars, we first estimated the mass of these objects and then used a parametric model-based analysis to derive the masses of the primary companion stars. The calculation is extended to 205 barium stars taken from literature. An analysis of the mass distribution revealed that both the strong and the mild barium stars occupy  the same range of masses. However, their companion AGB stars' mass distributions peak at two different values 2.5 M$_{\odot}$ and 3.7 M$_{\odot}$, for  the strong and the mild  barium stars respectively. This provides clear evidence that the formation of mild and strong Ba stars is greatly influenced by the initial masses of the companion AGB stars. It remains, however, to be seen the possible impact of other factors such as orbital period and metallicity on the formation scenarios of  barium  stars.
\end{abstract}

\keywords{Barium stars,  AGB, mass-distribution}

\section{Introduction}

\citet{Bidelman_Keenan_1951} first recognised abnormally strong Ba~II (4554 \AA) and Sr~II (4077 \AA) features in some giant stars and categorised these objects as ``Barium Stars''. In addition to these features barium stars are also characterised by enhanced abundances of slow neutron-capture process elements \citep{Bidelman_Keenan_1951, Warner_1965}. Several authors \citep{Bond_1984, Luck_&_Bond_1991} have suggested that the CH strong subgiants are likely progenitors of  barium stars as they exhibit similar carbon and s-process elements. With the s-process signatures of AGB stars similar to CH stars, the barium stars are also  believed to be the metal-rich Population~I analogues of CH stars \citep{Allen_Barbuy_2006}. From a long-term radial velocity observations of a large number of barium stars \citet{McClure_et_al_1980} have confirmed the binary nature of barium stars. \citet{Jorissen_&_van_2000} have suggested that the abundances of heavy elements observed in barium stars are the result of a mass transfer process and can be explained with the help of a binary picture including low-mass AGB  star.  Based on the abundances of neutron-capture elements barium stars are divided into two groups, the strong barium stars and the mild barium stars. However, the characteristic properties of the companion stars that are likely to play a vital role in making a barium star ``strong'' or ``mild'' are not clearly  understood. In this work, we have addressed this problem by considering  the mass distribution of the companion stars of  a large set of barium stars. In a recent work,  from  a detailed analysis of seven stars (BD+75348, BD+093019, HD238020, HE0319--0215, HE0507--1653, HE0930--0018, and HE1023--1504) based on  high-resolution spectroscopic analysis we  have identified two strong barium stars (BD+75348, BD+093019) and one mild barium star (HD238020) in the sample. These objects are included in the present study. In Section~\ref{sec:rv}, we present a brief discussion on  the radial velocities, stellar atmospheric parameters and elemental abundances of the stars. Section~\ref{sec:discussion} briefly discusses the determination of masses of the programme stars as well as  the masses of their primary companion stars. A discussion on the mass distribution and clues  derived from the distribution patterns are also presented in this section. The conclusions are drawn  in section~\ref{sec:conclusion}.

\section{Radial Velocities, Stellar Atmospheric Parameters and Elemental Abundances}
\label{sec:rv}

For our analysis we have used High-resolution ($\lambda/\delta \lambda$ $\sim$ 60~000) spectra acquired using Hanle Echelle SPectrograph (HESP) for the objects  BD+093019, BD+75348, and HD238020. High-resolution ($\lambda/\delta \lambda$ = 50~000) spectra of the other stars were taken from the \href{http://jvo.nao.ac.jp/portal}{\it{SUBARU archive}}. IRAF was used for data reduction and continuum fitting.
We have used broadband colours, optical, and infrared (IR) to determine the photometric temperatures of the programme stars. The colour-temperature calibrations used in this study are based on the IR flux method \citep{Alonso1996, alonso1999effective}. The T$_{eff}$(J--K) is independent of metallicity and hence we have used this temperature as an initial guess to select the model atmospheres by an iterative method. The radial velocities of the objects are calculated by measuring the shifts of the elemental absorption lines from the laboratory wavelengths, using Doppler's formula. The noticeable discrepancies between the radial velocities obtained by us and the values reported by \citet{gaia2018} suggest the possibility of the programme stars being  binary systems with  unseen companions. Stellar atmospheric parameters such as T$_{eff}$, log~$g$, microturbulent velocity ($\xi$), and [Fe/H] have been derived using the MOOG code \citep{Sneden_1973_MOOG} and the method described in \citet{Goswami_et_al_1_2021, Goswami_Goswami_2022}. The model atmospheres have been selected from the Kurucz grid of model atmospheres. Using MOOG code, we determined the abundances of the light and neutron-capture elements of the programme stars. Spectrum synthesis calculations were used for deriving the abundances of C, N, O and the elements showing hyperfine splitting. The equivalent width method was used for the rest of the elements. \href{https://lweb.cfa.harvard.edu/amp/ampdata/kurucz23/sekur.html}{\it{Kurucz database}} and \href{https://github.com/vmplacco/linemake}{\it{linemake}} provided the atomic line information.

The details of the methodology, linelists and the results of this study can be found in \citet{Goswami_&_Goswami_2023}.

\section{Results and Discussion}
\label{sec:discussion}

\subsection{Estimation of masses  of Ba stars and mass distribution}
\label{sec:discussion1}
Our analysis shows that BD+75~348 and BD+09~3019 are strong Ba ([hs/Fe] $>$ 0.60) stars and HD~238020 a mild Ba star (0.17 $<$ [hs/Fe] $<$ 0.60). In addition to these objects, we have compiled the abundance data and other relevant data of a large sample of Ba stars from the literature to derive the mass distribution of these stars. We have estimated the masses of the Ba stars  by determining their position on the HR diagram and used Equation~\ref{eqn:luminosity} to calculate their luminosities (L).

\begin{equation}
\label{eqn:luminosity}
log(L/L_{\odot}) = 0.4(M_{bol\odot} - V - 5 - 5log(\pi) + A_{v} - BC)
\end{equation}

\begin{figure}
     \begin{center}
\centering
        \subfigure[]{%
            \label{fig:1a}
            \includegraphics[height=6.5cm,width=8.0cm]{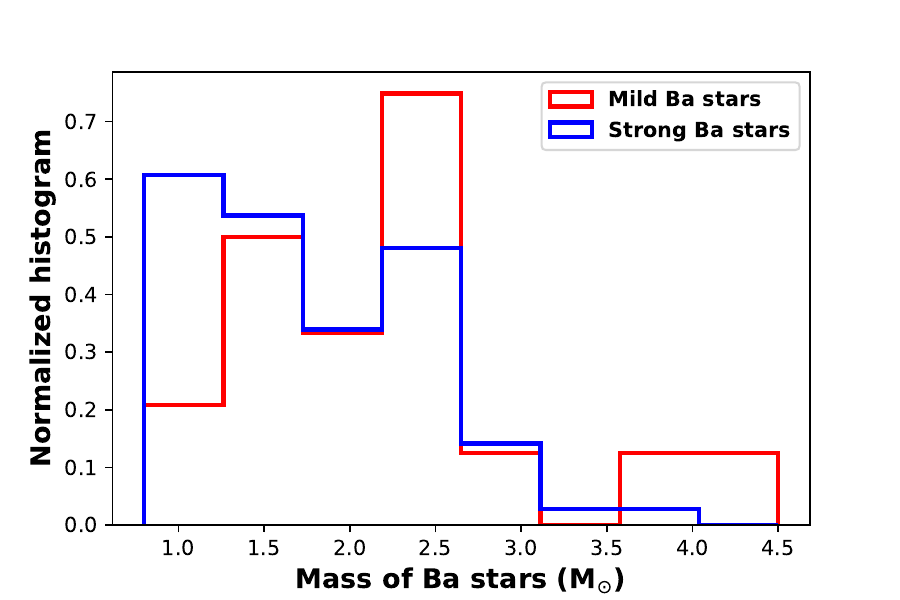}
        }%
        \subfigure[]{%
            \label{fig:1b}
            \includegraphics[height=6.5cm,width=8.0cm]{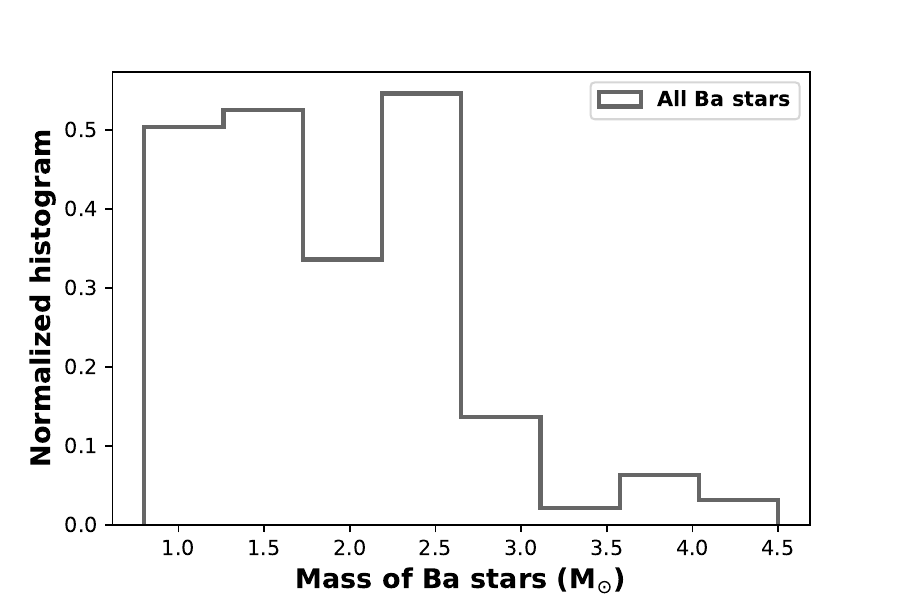}
        }\\ %  ------- End of the first row -------------
\bigskip
\begin{minipage}{12cm}
    \caption{Mass distributions of Ba stars. Panel (a) displays the mass distributions of mild and strong Ba stars separately, while Panel (b) shows the mass distribution of all mild and strong Ba stars together. Taken from \citet{Goswami_&_Goswami_2023} with permission. }%
   \label{fig:ba_primary_secondary}
\end{minipage}
       \end{center}

\end{figure}

We have used various sources to obtain the necessary data required for the analysis. The V magnitudes of the objects are obtained from SIMBAD, while the parallaxes ($\pi$) are sourced from \href{https://gea.esac.esa.int/archive/}{\it{Gaia}}. Interstellar extinction (A$_{v}$) values are calculated using the equation of \citet{Chen_et_al_1998}, and the bolometric corrections (BC) are determined using empirical calibrations by \citet{alonso1999effective}. To estimate the masses of the Ba stars, we used updated BASTI-IAC evolutionary tracks, corresponding to the metallicities of the Ba stars. In total, 205 Ba stars are analysed, including 52 mild and 153 strong Ba stars. We found that mild and strong Ba stars have similar mass ranges (0.8 to 4.5 M$_{\odot}$), with tail ends going to 4.5~$\pm$~0.25 M$_{\odot}$ and 4.0~$\pm$~0.25 M$_{\odot}$, respectively (Figure~\ref{fig:1a}). Additionally, the distribution of Ba stars does not follow a single Gaussian distribution but is scattered throughout the mass range. On average, the mass of Ba stars is estimated to be 1.9 M$_{\odot}$ (Figure~\ref{fig:1b}). Contrary to \citet{Escorza_et_al_2017} that reported a shortage of Ba stars in the 1.0 -- 2.0 M$_{\odot}$ mass range, our study shows that this range is significantly populated by Ba stars, as shown in Figure~\ref{fig:1a}.

\subsection{Estimation of masses of AGB companions and mass distribution }
\label{sec:discussion2}

We have used AGB yields of the FRUITY models \citep{Cristallo_et_al_2015} to perform a parametric-model-based analysis to estimate the initial masses of the companions of the programme stars and the sample of Ba stars compiled from the literature. The FRUITY models are used to calculate the s-process yields for various masses that correspond to the metallicities of the stars. Next, the observed abundances of the neutron-capture elements are compared to the model function to identify the best fit. The required mass of the companion AGB of the programme star is determined as the corresponding mass of the AGB model for which the minimum value of $\chi^{2}$ is achieved. The detailed methodology followed for the parametric-model-based study is same as that explained at length in \citet{Goswami_&_Goswami_2023}. We have determined the masses of the primary companions for 158 Ba stars, including 52 mild Ba stars and 106 strong Ba stars. 

\begin{figure}
     \begin{center}
\centering
        \subfigure[]{%
            \label{fig:2a}
            \includegraphics[height=6.5cm,width=8.0cm]{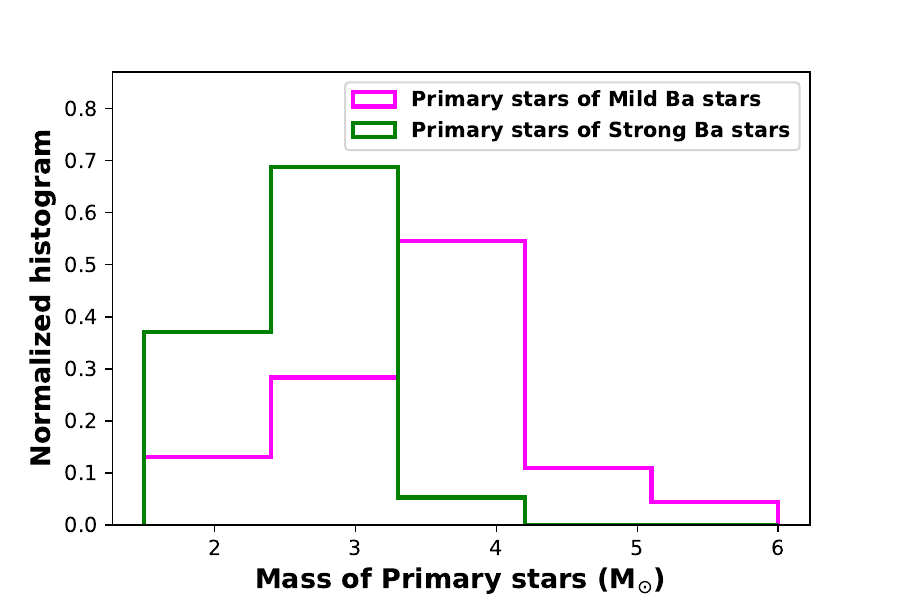}
        }%
        \subfigure[]{%
            \label{fig:2b}
            \includegraphics[height=6.5cm,width=8.0cm]{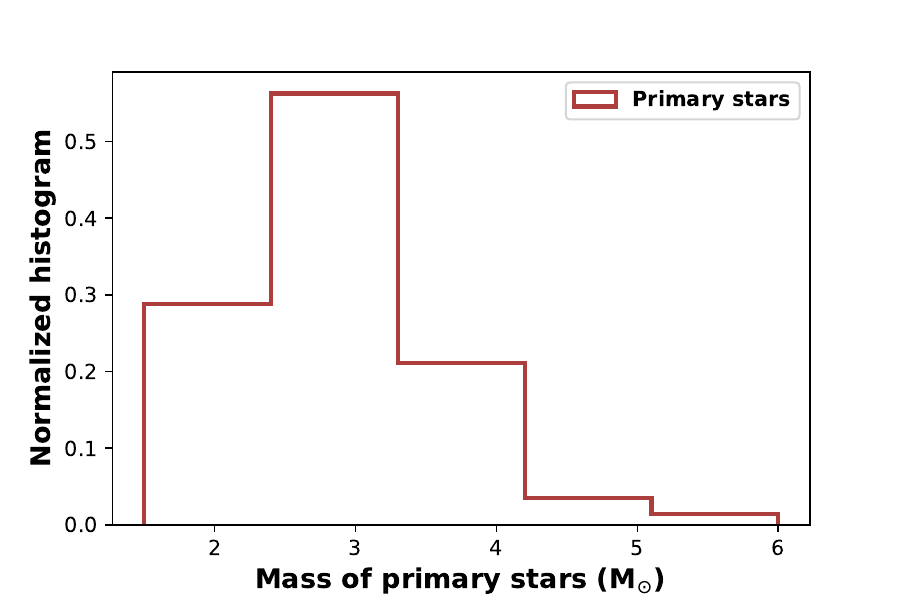}
        }\\ %  ------- End of the first row -------------

\bigskip
\begin{minipage}{12cm} 
    \caption{Mass distributions of the primary stars (progenitor AGBs) of Ba stars. Panel (a) displays the mass distributions of the progenitors of mild and strong Ba stars separately, while Panel (b) presents the mass distribution of the progenitors of both mild and strong Ba stars together.  Taken from \citet{Goswami_&_Goswami_2023} with permission.}%
   \label{fig:ba_primary_secondary2}
\end{minipage}
       \end{center}

\end{figure}

The mass distribution of the primary companions for mild and strong Ba stars is found to have two peaks with a larger  mass range for mild Ba stars. The distribution of primary companion masses of strong Ba stars peaks at 2.5 M$_{\odot}$, while that of mild Ba stars peaks at 3.7 M$_{\odot}$ (Figure~\ref{fig:2a}). The distribution of primary companion masses of Ba stars, when considered as a whole, peaks at 2.9 M$_{\odot}$ (Figure~\ref{fig:2b}).

\subsection{Derived clues on the formation Scenarios of the two classes of Ba Stars}
\label{sec:discussion3}
Theoretical analysis by \citet{Han_et_al_1995} indicates that most mild Ba stars are formed through wind accretion and wind exposure, whereas strong Ba stars are formed through wind accretion, wind exposure, and stable Roche lobe overflow (RLOF).
According to \citet{de_Castro_et_al_2016}, mildly enhanced interstellar matter with s-process elements is the source of mild Ba stars. However, subsequent studies by \citet{Jorissen_et_al_2019} have revealed that both mild and strong Ba stars are formed through AGB mass transfer in binary systems, rendering the previous scenario invalid.  \citet{Yang_et_al_2016} proposed two possible formation scenarios for mild Ba stars, but observations by \citet{Jorissen_et_al_1998,Jorissen_et_al_2019} have shown that mild Ba stars are not restricted to long-period systems and that the orbital periods of mild and strong Ba stars overlap. 

The differences in the peaks of primary masses for mild and strong Ba stars indicate that the companions of mild Ba stars are more massive. This implies that the initial mass of the companion is the main factor controlling the abundance peculiarities of mild Ba stars, with mass transfer from 4.0--6.0 M$_{\odot}$ AGB companions capable of explaining their formation \citep{Goswami_&_Goswami_2023}.
 White dwarf masses of approximately 1.0 M$_{\odot}$ around Ba stars also indicate a massive (5 M$_{\odot}$) companion AGB \citep{Jorissen_et_al_2019}. However, in the case of primary masses of mild and strong Ba stars in the range of 1.3--4.0 M$_{\odot}$, the system's metallicity and dilution are essential in their formation \citep{Goswami_&_Goswami_2023}. Mild Ba stars have a higher metallicity distribution than strong Ba stars, and the s-process efficiency decreases at higher metallicities, resulting in lower AGB yields.

\section{Conclusion}
\label{sec:conclusion}

We have investigated  the mass distribution of Ba stars and found that it cannot be represented by a single Gaussian. The average mass of the distribution is found to be 1.9 M$_{\odot}$ with strong and mild Ba stars extending up to 4.0 M$_{\odot}$ and 4.5 M$_{\odot}$, respectively. We confirmed that mild and strong Ba stars occupy a similar mass range. A previous claim  (i.e., \citet{Escorza_et_al_2017})  that there is a lack of Ba stars in the mass range of 1.0--2.0 M$_{\odot}$ is not supported by our study. Our investigation shows  that the progenitor mass distributions of mild and strong Ba stars peak at different values. We, therefore, suggest that the initial mass of the companion AGB is the primary factor in controlling the heavy elements' enrichment in  barium  stars, and thus plays a vital role in the making of strong and mild barium stars. However,  
the impact of orbital period and metallicity in the formation of mild Ba stars also need to be considered and investigated in detail, as there is significant overlap in the orbital periods of mild and strong Ba stars. Such a study has been undertaken and the results will be discussed in a separate paper.

\begin{acknowledgments}
This work made use of the SIMBAD astronomical database, operated at CDS, Strasbourg, France, the NASA ADS, USA and data from the European Space Agency (ESA) mission \href{https://www.cosmos.esa.int/gaia}{\it{Gaia}}, processed by the Gaia Data Processing and Analysis Consortium (DPAC, \url{https://www.cosmos.esa.int/web/gaia/dpac/consortium}).    AG  acknowledges  the support received from
the Belgo-Indian Network for Astronomy \& Astrophysics  project BINA - 2  (DST/INT/Belg/P-02 (India) and BL/11/IN07 (Belgium)).    
\end{acknowledgments}

\begin{furtherinformation}

\begin{authorcontributions}
PPG  and AG  designed the research; PPG compiled the data and performed the research; PPG and AG interpreted the results and  wrote the paper.

\end{authorcontributions}

\begin{conflictsofinterest}
``The authors declare no conflict of interest.''
\end{conflictsofinterest}

\end{furtherinformation}

\bibliographystyle{bullsrsl-en}

\bibliography{sample}

\end{document}